1# A New Intelligent Cross-Domain Routing Method in SDN Based on a Proposed Multiagent Reinforcement Learning Algorithm

Miao Ye, Linqiang Huang, Xiaofang Deng, Yong Wang, Qiuxiang Jiang, Hongbing Qiu, Peng Wen

*Abstract*—Message transmission and message synchronization for multicontroller interdomain routing in software-defined networking (SDN) have long adaptation times and slow convergence speeds, coupled with the shortcomings of traditional interdomain routing methods, such as cumbersome configuration and inflexible acquisition of network state information. These drawbacks make it difficult to obtain a global state information of the network, and the optimal routing decision cannot be made in real time, affecting network performance. This paper proposes a cross-domain intelligent SDN routing method based on a proposed multiagent deep reinforcement learning. First, the network is divided into multiple subdomains managed by multiple local controllers, and the state information of each subdomain is flexibly obtained by the designed SDN multithreaded network measurement mechanism. Then, a cooperative communication module is designed to realize message transmission and message synchronization between root and local controllers, and socket technology is used to ensure the reliability and stability of message transmission between multiple controllers to realize the real-time acquisition of global network state information. Finally, after the optimal intradomain and interdomain routing paths are adaptively generated by the agents in the root and local controllers, a prediction mechanism for the network traffic state is designed to improve the awareness of the cross-domain intelligent routing method and enable the generation of the optimal routing paths in the global network in real time. Experimental results show that the proposed cross-domain intelligent routing method can significantly improve the network throughput, reduce the network delay and the packet loss rate compared to the Dijkstra and OSPF routing methods.

This work was supported in part by the National Natural Science Foundation of China (Nos.62161006, 62172095, 61861013), the subsidization of the Innovation Project of Guangxi Graduate Education (No. YCSW2022271), and Guangxi Key Laboratory of Wireless Wideband Communication and Signal Processing (No. GXKL06220110). (Corresponding author: Xiaofang Deng).
Miao Ye, Xiaofang Deng, Qiuxiang Jiang, Hongbing Qiu and Peng Wen are with School of Information and Communication, Guilin University of Electronic Technology,Guilin ,541004,China (email:yemiao@guet.edu.cn; xfdeng@guet.edu.cn; jiangqiuxiang@guet.edu.cn; qiuhb@guet.edu.cn; wenpeng1994325@gmail.com )

Linqiang Huang and Yong Wang are with School of Computer and Information Security, Guilin University of Electronic Technology, Guilin, 541004, China (e-mail: huanglinqiang_2020@foxmail.com; ywang@guet.edu.cn).
## I. INTRODUCTION

In recent years, with the development of new trendy technologies such as 5G networking, and cloud computing, various multimedia services and network devices have undergone rapid development. These network devices usually have fast response times, high network performance, and multiple functional types to ensure efficient quality of service (QoS) for users. The most basic and vital condition for meeting QoS requirements is to ensure the network performance quality, which requires an efficient QoS-aware network architecture. Software-defined network (SDN) [1][2], as a new network architecture, can separate the control plane and data plane. Network state information is obtained according to the SDN southbound interface, and the SDN northbound interface provides upper-layer application services. In this way, centralized and unified management of the network can be realized, and a global state information of the network can be easily obtained to support the flexible deployment of routing strategies, thereby improving network performance. Therefore, SDN is considered a very effective QoS-aware network architecture.

With the SDN characteristics of open programming and separation of data plane and control plane architecture, many researchers have used SDN to implement routing and forwarding and traffic management in data center networks (DCNs) and traffic engineering (TE) and have achieved good results. However, as the network scale increases, SDN based on management by means of a single controller will encounter problems such as excessive controller load and single point failure; moreover, the accumulation of traffic packets will lead to long queue wait times and a lack of timely forwarding, which will seriously affect network performance. To solve the bottleneck problem for large-scale networks under the single-controller management mode, domainwise management based on the multicontroller mode has become a key technology. In this approach, a large-scale network is divided into multiple network subdomains, a so-called local controller is deployed in each subdomain for traffic management and routing/forwarding in that local domain, and a root controller is introduced with the function of coordinating the global network for cross-domain information exchange and routing/forwarding. Thus, the problems of excessive load and data packet accumulation in a large-scale network under the single-controller SDN management mode can be solved.

However, the critical problems of message transmission and message synchronization between controllers have yet to be solved for multicontroller domain management in large-scale networks. The traditional Border Gateway Protocol (BGP) [3] transmits messages between adjacent autonomous systems (ASs) through border routers to achieve message synchronization between domains. However, the configuration of message transmission and message synchronization based on BGP in an SDN environment is cumbersome and suffers from problems such as routing oscillation. The OpenFlow1.3 protocol [4] provides an east–west interface to realize the delivery of cross-domain SDN routing messages. However, in contrast to the widespread use and standardization of north–south interfaces, no consensus standard for east–west interface has yet been formed in the industry. Some scholars have made achievements in work on east–west interfaces for SDN [5][6][7], through which consistency of the global network state can be achieved via adaptive updating of the state between controllers, effectively solving the problems of message transmission and message synchronization between multiple controllers. However, these adaptive methods for achieving message transmission and message synchronization based on multiple controllers have problems such as long adaptation times and slow convergence speeds and do not permit the entire network view state information to be obtained in real time. In addition, ensuring the reliability and stability of message transmission between multiple controllers is vital to ensure the performance of the SDN multi-agent routing optimization method. To solve these problems, this paper proposes a cross-domain intelligent SDN routing method based on a proposed multiagent deep reinforcement learning algorithm (MDRL-TP), SDN multithreaded measurement technology is applied to obtain the link state information in each subdomain of the network, and socket [8] technology is used in a cooperative communication module for point-to-point communication between multiple controllers to achieve interdomain message transmission and message synchronization, thereby improving the convergence speed of obtaining the global network state information and ensuring the reliable and stable transmission of messages between multiple controllers. Then, through the high-dimensional feature processing ability of deep learning (DL) [9], the exploration and exploitation capabilities of reinforcement learning (RL) [10] for decision-making, and the time-series prediction capability of recurrent neural networks (RNNs) [11] After iterative training, the optimal intradomain routing forwarding paths between all source-destination switches in each network subdomain can be generated in real time, and the optimal interdomain routing forwarding paths can be generated according to the interdomain routing forwarding request of the network, so as to obtain the optimal routing forwarding path of the whole network.

The main innovations are as follows:

1) Design a fast and stable module for obtaining global network state information using the multithreaded network measurement mechanism of SDN. Based on the multithreaded network measurement mechanism of SDN, the link state information in each network subdomain is flexibly obtained, and the designed cooperative communication module uses socket technology for point-to-point communication between the root and local controllers to realize interdomain message transmission and message synchronization, thereby improving the convergence speed of obtaining global network state information and ensuring the reliability and stability of message transmission between multiple controllers.

2) Propose a network traffic state prediction model in the SDN cross-domain intelligent routing method under the multiagent deep reinforcement learning mechanism. By adopting gated recurrent unit (GRU), a network traffic state prediction model is designed in paper, which is used to monitor the hidden traffic states in the network under multicontroller management to improve the perception and performance of the cross-domain intelligent routing method. As a result, the proposed cross-domain intelligent SDN routing algorithm is easier to converge and more concise.

3) Develop a multiagent reinforcement learning algorithm that generates optimal route forwarding paths in large-scale networks in real-time. The hierarchical architecture is used to divide a large-scale SDN network into multiple subdomains, and multiple SDN agents are trained concurrently through distributed technology to reduce the time overhead for model training. After the agents in the local and root controllers adaptively generate the optimal intradomain and interdomain routing forwarding paths under the network state at each moment, the MDRL-TP multiagent cross-domain routing algorithm generates the optimal routing forwarding paths for the global network in real time, thus effectively solving the problems of excessive load and traffic packet accumulation in a large-scale network under the management of a single SDN controller and realizing the ability to make real-time intelligent optimal routing decisions in large-scale networks.

The remainder of the paper is as follows: Section II describes the current state of research and problems of SDN multicontroller architectures, multi-intelligence deep reinforcement learning mechanisms, and multicontroller routing algorithms. Section III describes in detail the multiagent cross-domain intelligent routing optimization architecture and modeling based on the SDN. Section IV introduces the proposed multiagent cross-domain intelligent SDN routing method and details the specific algorithm design. Section V describes the experimental environment and performs experimental analysis and validation of the MDRL-TP multi-intelligent body routing method. Section VI gives conclusions and describes the directions for further research in the future.

## II. RELATED WORK

With the continuous development of artificial intelligence technology, deep reinforcement learning (DRL) algorithms have shown significant advantages in solving complex decision-making problems. Many researchers have applied DRL to SDN routing optimization and achieved good results.

Dai et al. [12] proposed an algorithm for link state estimation (IQoR-LSE) to meet the diverse service quality requirements of various types of network applications. Under the link congestion inference conditions, this algorithm guides action space exploration in DRL to find the optimal routing strategy and solves the problem of nonconvergence in the exploration of complex action space through the joint estimation of link congestion, which significantly reduces the jitter and packet loss rate of the network. To cope with the exponential growth of network traffic demand, Kim et al. [13] designed a DRL-based SDN routing algorithm. By learning the correlations between traffic balance and network load of SDN switches through DRL, a set of optimal link weights was generated for routing and forwarding. Meanwhile, to reduce the training time of DRL, a method model based on an M/M/1/K queue was designed to improve the algorithm's robustness. Sun et al. [14] proposed the ScaleDeep route optimization method with SDN scalability, aiming to address the fact that a routing optimization algorithm based on DRL usually relies on information from all nodes in the network for optimal decision-making, resulting in problems such as slow convergence and susceptibility to interference from network topology changes in large networks. Based on control theory, a group of key nodes is selected as the driver nodes in the network, and the link weights are dynamically adjusted by monitoring the traffic changes at the driver nodes, thereby improving the routing performance and reducing the influence of network topology changes on network performance. Xia et al. [15] proposed a QoS optimization method for heterogeneous networks, taking both network delay and load balance as objective functions based on the consideration that routing algorithms based on single metric parameters have difficulty meeting the QoS of multisource data flows in heterogeneous networks, leading to problems such as link congestion and network resource waste. The Double Deep Q-Network (DDQN) algorithm can generate the optimal paths for multisource data flows, thus improving load balancing and routing efficiency in heterogeneous networks. Considering that the dynamic complexity of a network and the future change trend of the network traffic state will affect the network performance, our previous work [16] proposed an intelligent SDN routing method based on Dueling Deep Q-Network (DQN) DRL and network traffic state prediction (DRL-TP). In this method, a GRU prediction algorithm is used to improve the perception ability of the DRL, realize real-time intelligent routing decision, and effectively improve the performance of the network. However, the above intelligent routing methods are all based on single SDN controllers. When the network scale continues to expand, more and various types of data flows will be present in the network, resulting in severe load and data accumulation for a single SDN controller, which will make it impossible to make real-time intelligent routing decisions. Therefore, intelligent routing decision-making methods based on single-controller SDN management cannot meet the actual requirements of large-scale networks, and it is essential to design a distributed network architecture that can allow large-scale networks to be divided into multiple subdomains for management. A multicontroller architecture for SDN is considered an effective solution to this problem.

At present, most research work [17][18][19][20] mainly divides SDN multicontroller architectures into two forms: a flat architecture and a hierarchical architecture. In the flat architecture, each controller is at the same level, and neighboring controllers communicate to exchange information and thus obtain global network link state information. The hierarchical architecture consists of at least two layers, the first consisting of the root controller and the second consisting of the local controllers. Each local controller is used for information interaction and routing decisions in its local domain, while the root controller cooperates with the other controllers and is used for interdomain information interaction and routing decisions. Although the flat architecture is an extension functionality of the SDN control plane, it requires more complex collaborative functions and control overhead. For example, frequent communication between controllers is required to keep the global network view consistent; the change of network topology under a controller will affect the acquisition of global view information. In contrast, the hierarchical architecture can simultaneously obtain network link information managed by multiple controllers, and thus, the global network link information can be obtained more quickly. Considering that the purpose of this paper is to solve the problem of intelligent routing optimization in large-scale networks under SDN multicontrollers, which requires the real time acquisition of global network link information, the hierarchical architecture is chosen based on the characteristics of the two architectures.

Routing optimization based on multiple SDN controllers has always been a popular research topic. Many researchers have reported achievements with respect to this problem using traditional modeling optimization methods and heuristic algorithms. Wang et al. [21] considered that the distributed controller cluster in dealing with large-scale SDN will bring the local controller to the root controller to report the interdomain flow to the root controller, resulting in response time overhead and link load imbalance. A rounding-based algorithm was proposed to optimize the SDN controller load, effectively reduce the processing response delay of the controller, and achieve the balance of the link load. However, this method requires the construction of relevant models and optimization goals, and it is challenging to build a model that conforms to the multiple network subdomains in a large-scale network managed by multiple SDN controllers. Moufakir et al. [22] proposed a new multidomain cooperative routing framework. First, a proxy-based routing framework was proposed to ensure the performance and cost of cross-domain network data flows. Then, a greedy algorithm was designed to handle large-scale network problems with multiple domains so as to maximize the overall network utilization rate and reduce the network delay cost. However, this greedy algorithm cannot be guaranteed to obtain the globally optimal solution. Xue et al. [23] proposed a compact evolutionary tabu algorithm (CETS) to solve the shortcomings of traditional evolutionary algorithms in finding global optimal solutions in sensor networks, such as premature convergence and long search time, which effectively improves the matching efficiency of sensor networks. To achieve efficient routing and low energy consumption in a distributed SDN architecture, EL-Garoui et al. [24] proposed an energy-aware routing multilevel mapping algorithm (EARMLP), which seeks the optimal routing solution between controllers and between controllers and switches by means of an energy-aware consumption strategy, thus improving network performance. This method also uses a correlation model to find the optimal solution. However, it is still challenging to construct multiple network subdomain models for large-scale networks. Xu et al. [25] designed an improved multi-objective optimization genetic algorithm (GAIMO) to address the problem of unbalanced controller load due to multiple types of network traffic in SDN multi-controller managed networks, using the network latency and switch transfer cost as decision factors. Aiming at the problem that a genetic algorithm can easily fall into a locally optimal solution and thus fail to find the globally optimal solution, a local search operator was designed to reduce the convergence time of the algorithm and enhance the performance efficiency of the algorithm. The algorithm reduces the resource utilization of the SDN controller and optimizes the performance of the whole network. However, this method is subject to strict requirements in its application scenarios. Changes in the network topology and in intradomain and interdomain links will reduce the stability and performance of the heuristic method, thereby affecting the performance of the entire network. Zhang et al. [26] designed an adaptive synchronization strategy to establish state synchronization among multiple SDN controllers to ensure the consistency of information, thus effectively reducing the communication overhead between controllers and improving network performance. However, when the network topology changes, this method needs to spend a long time for adaptive synchronization, which makes it challenging to meet the needs of a dynamic network. With the development of the internet, high requirements for end-to-end QoS (E2E QoS) assurance are emerging, especially for different types of traffic across domains; consequently, strong stability and robustness are required to coordinate and ensure E2E QoS. Podili et al. [27] proposed a trust-aware E2E QoS routing (TRAQR) architecture to provide trusted E2E QoS in multi-domain SDN (MD-SDN), effectively ensuring network quality and performance. Nevertheless, this method also faces a potential network scalability problem when the network topology changes, which will affect the performance of the whole network.

Artificial intelligence technology has enabled remarkable achievements in many fields in recent years and has gradually advanced from perceptual intelligence to decision-making intelligence. DRL is a critical method for realizing decision intelligence, and because multiagent interactions often exist in the real world, this situation has led to the development of multiagent deep reinforcement learning (MDRL) [28]. Many scholars have begun to study MDRL mechanisms and have reported many related achievements [29][30][31]. Currently, MDRL is mainly divided into centralized learning, independent learning, and centralized training with decentralized execution (CTDE).

In centralized learning, all agents are trained as a whole; that is, the state-action spaces of each agent are combined to generate union state-action spaces, which integrate the state information of all agents, making it easier to obtain the global optimal decision. Nevertheless, with the increasing number of agents, the dimensions of the state and action spaces will increase exponentially, and centralized learning cannot take advantage of distributed training, resulting in colossal exploration and trial-and-error time costs for agents and excessive time consumption for model training. In addition, under such a reinforcement learning mechanism based on multiple agents, the global state needs to be obtained in each round of training, which is infeasible in practical application scenarios.

Independent learning refers to the stacking of multiple independent agents. Each agent has its own environment and does not interfere with other agents. During the training process, each agent updates its network model independently and selects its own optimal strategy to maximize its own reward value. The advantage of this approach is that there is no need to consider cooperation or interaction among various agents, meaning that the implementation is relatively simple. Distributed technology can accelerate the training time for multiple agents, and this approach is suitable for problems with discrete states and small action spaces. However, the independent learning method requires that the agents exist in a relatively balanced environment. Once the environment is dynamically changing, the agents' stability and convergence are difficult to guarantee.

CTDE is a combination of centralized learning and independent learning that combines the advantages of the two approaches. It uses distributed technology to speed up the model training time for multiple agents while allowing the agents to use global state information to conduct centralized learning and training in accordance with corresponding requirements. After training, the multiple agents make a distributed optimal decision. CTDE is a common approach for multiagent reinforcement learning.

Taking advantage of multiagent RL in solving problems involving large-scale networks, Li et al. [32] designed a load algorithm based on RL to solve the local overload problem for multiple SDN controllers. By constructing triples (out-migration domain, in-migration domain, and switch to be migrated) under constraints of the best overall migration performance and no migration collision, the triple representing the final migration for the current round of iteration was selected based on reinforcement learning, and the globally optimal solution was found at the minimum cost, which is used to solve the problem of excessive load between controllers and reduce the migration costs, thus achieving the fast response of request data packets. However, this method faces challenges related to constructing triples, ensuring the best overall migration efficiency under the current triple, and shortening the time overhead of iteration. Considering that traditional routing uses some information to implement forwarding operations, which can not achieve rapid adaptation to traffic variability and ensure the QoS of the network, Casas-Velasco et al. [33] proposed a DRL and SDN intelligent routing (DRSIR) method that considers state metrics such as available bandwidth, latency and loss path priority as the best routing to generate efficient intelligent routing adapted to variable traffic, and using natural and synthetic traffic matrix to evaluate DRSIR through simulation, significantly reduce the loss of packet and reduce the network delay cost. However, this method does not consider the influence of the future trend in network traffic state on network performance. Wang et al. [34] proposed a collaborative flow management framework to improve the flow management performance in multidomain SDN networks by finding the minimum link weight path for each flow to balance the link loads, thereby effectively relieving congestion and significantly improving the network throughput. However, this method also does not consider the impact of the change trend of network traffic state on network performance in the future. Godfrey et al. [35] designed a hierarchical SDN architecture with multiple controllers and proposed a Q-routing hop weight adjustment method for wireless sensor networks. A super controller obtains network information from the local domain controllers, and multiple factors (sensor energy, congestion, and link quality) are considered when selecting the next hop, effectively improving the wireless sensor network performance. However, with the increase of network size, this method uses a Q-table to find the optimal routing strategy, and the query time overhead and storage space capacity will present significant challenges. Yuan et al. [36] designed a distributed cooperative DRL method based on hierarchical network architecture control structure, and they used this method to solve the problems of controller dynamic allocation and delay control in vehicular networks. Through distributed cooperative operation, local controllers and neighboring controllers coordinate local strategies, effectively reducing the control delay and packet loss. However, this method does not consider dynamic controller allocation, the performance of multicontroller cooperation, or network state consistency. Considering the time costs incurred by heuristic iterative network performance optimization algorithms, Sun et al. [37] designed a controller load dynamic balance method based on multiagent reinforcement learning (MARVEL), which can overcome network load imbalance given a static mapping between the controller-switches under network traffic fluctuations. Under this method, the SDN controller will reject new network traffic requests due to overload, thus reducing the processing capacity of the control plane. This method also does not consider the coordination among controllers or the consistency of the network state.

To solve the shortcomings of the above work, the paper proposes a cross-domain intelligent SDN routing method based on a proposed MDRL algorithm. A SDN multithreaded network measurement mechanism is designed to obtain the traffic matrix from each network subdomain in real time, and a cooperative communication module is adopted that uses socket technology for point-to-point communication between multiple controllers to realize interdomain message transmission and message synchronization, thereby improving the convergence speed of obtaining global network state information and ensuring the reliability and stability of message transmission between multicontrollers, and then the global traffic matrix of the network can be obtained in real time. The long short-term memory (LSTM) [38] variant GRU [39] is used as the basis for a prediction algorithm to monitor hidden network traffic states in order to improve the awareness of the multiagent routing method, and finally, an action strategy is found by MDRL. This action strategy determines the globally optimal routing and forwarding paths given the current network state, and achieves adaptive intradomain and interdomain routing and forwarding operations.

## III. MULTIAGENT CROSS-DOMAIN SDN ROUTING OPTIMIZATION ARCHITECTURE AND MODELING

As shown in Fig. 1, the SDN multi-agent cross-domain routing optimization architecture in this paper is mainly consists of root controller, local controller and cooperative communication module. As the proposed work of multiagent cross-domain intelligent SDN routing method is the extension of our previous work which is only research the routing strategy in one SDN controller domain, here we also focused the same SDN architecture, so the data plane, control plane, management plane, and knowledge plane are deployed on multiple SDN controllers. Due to space limitations, only the newly added and improved functions on these four planes will be introduced in detail in this paper. For other similar functions, please refer to [16].

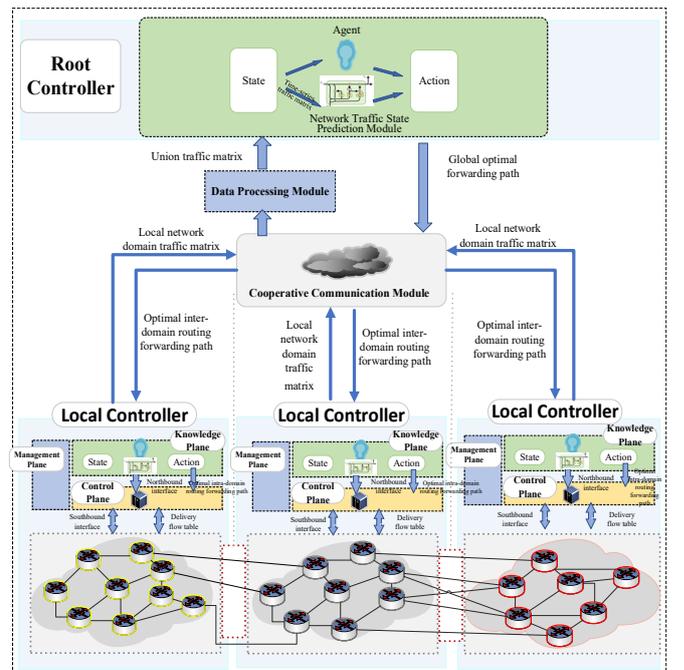

Fig. 1 Multiagent cross-domain SDN routing optimization architecture

## A. Root Controller

As the center of the multiagent cross-domain SDN routing optimization architecture, the root controller is mainly applied to collecting the global link state information of the network and generating the optimal routing and forwarding paths between domains. Therefore, the *management plane* and *knowledge plane* are deployed on the root controller.

*Management Plane:* The management plane provides the following functions: 1) Storing the collected link state information from multiple network subdomains in an information pool. 2) Converting the network link information saved in the information pool into a union traffic matrix (*UTM*) to be provided to the knowledge plane for offline model training.

*Knowledge Plane:* The knowledge plane consists of two modules: DRL and network traffic state prediction. The primary functions of the DRL module, that is, the agent, mainly uses the global traffic matrix provided by the management plane to build a network training environment, and continuously interacts with the built environment through the DRL model with high-dimensional decision-making in order to gain higher rewards. When the model training process converges, the action given the current network state is determined, i.e., the optimal routing and forwarding paths.

To make the optimal forwarding strategy, the knowledge plane needs to rely on the SDN controller in each network subdomain to continuously and frequently process the data flow in the network, and then obtain the global network state information in real time. Nevertheless, this response processing mechanism will make the SDN controller overload and consume too much, and there is a certain risk of downtime, resulting in some data flows in the network not being processed in time, resulting in network fluctuations, and affecting the normal operation of the SDN. Thereby, a multithreaded SDN network measurement mechanism is designed to effectively solves the problem of high load and high consumption caused by the SDN controller in each network subdomain when continuously processing high-volume data streams, and also provides the union traffic matrix to the knowledge plane in real time. However, there will still be omissions in the monitoring of some intra-domain traffic matrices in each network subdomain. In order to solve this problem of missing monitoring data in the intervalwise SDN network measurement mechanism, a network traffic state prediction module is deployed in the knowledge plane for predicting omissions and monitoring hidden intradomain traffic matrices.

In this paper, the hierarchical architecture is used to build a multicontroller SDN network that can simultaneously collect link state information for the subdomains under the management of each local controller and collect network global link information faster. However, some problems still remain, such as the excessive load on the root controller caused by large-scale network traffic requests. The solution is to use mechanisms such as root controller expansion or traffic diversion, which have enormous hardware resource requirements. Considering the experimental conditions and practical application scenarios, this paper adopts an intervalwise request response mechanism to solve the problem of excessive load on the root controller. Specifically, the optimal intradomain routing forwarding paths in each network subdomain are generated based on the routing decisions of the knowledge plane in the local controller of each subdomain, and the optimal interdomain routing forwarding paths are generated based on the routing decisions of the knowledge plane in the root controller at the request of the local controllers. This effectively solves the problem of the excessive load on the root controller caused by the need to make routing decisions for many network traffic requests in a large-scale network. However, considering that DRL requires continuous exploration and trial-and-error to ensure that the actions taken by agents are optimal, in addition to passively responding to interdomain requests for routing decisions, the root controller will also actively acquire the network global state information at intervals for learning to ensure that the corresponding agent can generate optimal routing decisions.

## B. Local Controllers

In this paper, the whole network under consideration is divided into three subdomains, and each local controller manages one subdomain. The main functions of a local controller are: 1) Processing the routing and forwarding requests in its local domain and generating the optimal intradomain routing forwarding paths. 2) Sending interdomain routing requests to the root controller and generating the optimal interdomain routing forwarding strategy after receiving the optimal interdomain routing forwarding paths from the root controller in response. The *data plane*, *control plane*, management plane, and knowledge plane are all deployed in each network subdomain.

*Data Plane:* The data plane provides the following functions: 1) Providing global awareness information in the local network domain, such as the switch port rate, the number of sending packets and bytes received, by responding to corresponding requests sent periodically by the control plane .2) After the agent generates the optimal intradomain routing forwarding path into the control plane to generate the optimal routing forwarding strategy, the data packet is forwarded by the data plane.

*Control Plane:* The control plane mainly deploys network information awareness module and the routing forwarding module. The network information awareness module consists of two functions: 1) Using the Link Layer Discovery Protocol (LLDP) to collect the topology information in the network. 2) The detailed state information of all switches in each network subdomain is periodically queried by the request instructions, and the awareness information in this domain is generated. The routing forwarding module is mainly divided into intradomain and interdomain forwarding. The main functions of intradomain forwarding include: 1) Finding corresponding host nodes in accordance with the network topology of the local domain based on the optimal intradomain routing and forwarding paths obtained from the knowledge plane in the local controller. 2) Generating an optimal intradomain routing policy based on the optimal intradomain routing and forwarding paths and corresponding host nodes and delivering the policy to the data plane. The main functions of interdomain forwarding include: 1) Sending interdomain routing forwarding requests to the root controller. 2) Receiving the optimal interdomain routing and forwarding paths generated by the knowledge plane of the root controller in response. 3) Generating the optimal interdomain routing and forwarding policy based on the optimal interdomain routing and forwarding paths and corresponding host nodes and delivering it to the data plane.

*C. Cooperative Communication Module*

At present, SDN east–west interface and BGP are commonly used for communication and collaboration among multiple network subdomains. The root controller requires a reliable mechanism for data transmission in order to obtain the network state from each local controller and thus obtain the global network state information in real time. However, east–west interface, which is still in the initial research stage, has difficulty meeting such requirements. BGP is a decentralized routing protocol that maintain routing information between different ASs for large-scale data centers. It is mainly used to exchange reachable routing information between ASs, construct interdomain propagation paths, and realize communication and routing decisions among multiple network domains. It is widely used for making routing decisions between different domains in traditional networks, but a problem of routing oscillation arises when routing strategies change. Moreover, as the network scale increases, the more routing entries there are, and the more complex the configuration and management problems become.

In this paper, a collaborative communication module is designed using socket technology, which is used to realize collaboration and message communication among multiple SDN domains. A socket is simply an endpoint abstraction of two-way communication between application processes on different hosts in a network. It is the basic operation unit of the TCP/IP communication protocol. Therefore, the socket mechanism is a reliable transport mechanism. The socket communication process is straightforward. Through the server binding IP address and port connection, the client realizes the communication between the client and the server and the network information transmission. This paper uses the root controller as the server and the local controller as the client. The message bodies between them are encapsulated in the *Response* and *Request* data structures and transmitted through the JSON format. The main functions of the cooperative communication module include the following: 1) A local controller sends an optimal interdomain routing forwarding path request to the root controller through a socket connection. 2) The root controller responds to the request and transmits the optimal interdomain routing forwarding path information to the local controller through the socket technology. 3) The root controller actively obtains the traffic matrix of each network subdomain through socket connections to construct the *UTM*.

The performance of the multiagent SDN routing algorithm is determined by whether the cooperative communication module can quickly and efficiently transmit requests from the local controllers to the root controller, responses from the root controller to the local controllers, and the traffic matrix of each network subdomain to the management plane. Therefore, multithreading and pipeline technology are used in this paper to ensure that the sockets in the collaborative communication module can efficiently process requests and responses and obtain the traffic matrix of each network subdomain. Specifically, three threads are created for each network subdomain: one to handle the local controller's request, one for the root controller's response, and one for the acquisition of the traffic matrix. Thus, in each network subdomain managed by the local controller, the network traffic matrix of the domain is obtained using the SDN multithreaded network measurement mechanism, as shown in Fig. 2. The primary process is to create an information detection thread, which will send a request to the local-domain control plane to monitor the current network awareness information every $t$ seconds. After receiving the monitoring request, the local-domain control plane will send a Request instruction to return the obtained network awareness information. Then, the data operation thread converts the obtained network awareness information into network link information such as residual bandwidth and delay through the relevant equation and constructs the traffic matrix in the network domain. The data operation thread is the center of the multithreaded SDN network measurement mechanism. It performs two different functions according to the training and application phases of the intelligent routing method. The blue arrows in Fig. 2 is the training phase, which mainly stores the intradomain traffic matrix constructed at each moment to the matrix pool through the information operation thread for offline training. The red arrows in Fig. 2 is the application phases, which mainly inputs the traffic matrix into the trained model to generate the optimal intradomain routing forwarding path from the knowledge plane in the network domain, and then generates the optimal intradomain routing forwarding strategy through the control plane. Finally, evaluation metrics such as network throughput are calculated according to the relevant equation for performance comparison in this domain.

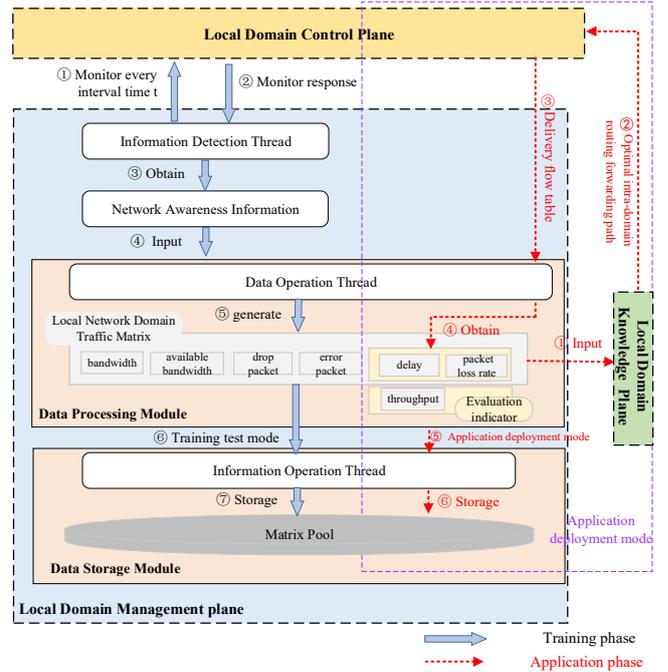

Fig. 2 Multithreaded SDN network measurement mechanism

Since sockets are a limited resource, frequent communication between the root and local controllers, especially for the active acquisition of the traffic matrix of each network subdomain by the root controller at regular intervals, will inevitably impose certain resource and time overheads. Therefore, to reduce the number of socket interactions when obtaining the traffic matrix from each network subdomain and increase the size of the data transmitted during each socket communication, pipeline technology is used, which is essentially a buffer. When the data size in the buffer meets certain conditions, the data in the buffer are transmitted in batches.

In this paper, the process of message communication and collaboration between the root and local controllers involves various operations, such as setting the name of each local controller, identifying which thread is processing a request, and identifying which thread is processing the response. Various status codes are defined to facilitate the above operations; the corresponding symbols and meanings are listed in Table I.

## IV. MDRL-TP Multiagent Cross-domain Routing Algorithm Design

The topology in the data plane of this paper is an undirected graph that can be represented by $G = \{D, V, E, W\}$, where $D$ is the current network subdomain, $V$ is the SDN switch nodes, $E$ is the link between SDN switches, and $W$ is the link weight size, which is generally set to a constant. Fig. 3 shows the flowchart of the MDRL-TP multiagent cross-domain routing algorithm.

First, the optimal intradomain forwarding paths of three network subdomains are generated by agents 1, 2, and 3 in the three local controllers. The specific process is as follows: 1) The traffic matrix in each subdomain is collected through the multithreaded SDN network measurement mechanism, the predicted traffic matrix is generated by the network traffic state prediction algorithm, which together constructs the state space of the DRL algorithm. 2) The DRL and prediction model in the local controller are trained based on offline mode, and the DRL-TP algorithm model in the current network subdomain is formed. 3) Collecting the traffic matrix in the network subdomain according to the current flow size, and generate the optimal intradomain routing and forwarding path in real time through DRL-TP algorithm.

Then, the agent in the root controller actively obtains the global network link information at regular intervals to train its model offline and passively responds to requests from each network subdomain to generate optimal interdomain routing forwarding paths. The specific process is as follows: 1) The traffic matrix from each network subdomain is obtained through the cooperative communication module, the union traffic matrix is obtained after the necessary data operations by the management plane, and then the union prediction traffic matrix is generated by the network traffic state prediction algorithm, which together constructs the union state space. 2) The DRL and prediction model in the root controller are trained through offline mode, and the DRL-TP algorithm model is formed. 3) According to the optimal interdomain routing forwarding request in each network subdomain, optimal interdomain routing forwarding paths are generated by the intelligent routing algorithm in the root controller.

TABLE I
STATUS CODE INFORMATION

| Status code | Symbol | Meaning |
|---|---|---|
| 0001 | SET_CONTR OLLER_NAME | Set local controller name |
| 0010 | ADD_DOMAI N_TOPO | Identify the network topology and add it to the corresponding local controller |
| 0011 | SYN_GLOBA L_VIEW | Identify the global link state information of the network |
| 0100 | ADD_INTER_ DPID | Confirm the interdomain switch in the current network to find the interdomain host |
| 0101 | REQ_INTER_ PROPERTY | Identify a request from a local controller for optimal interdomain routing forwarding paths |
| 0110 | RES_INTER_ PROPERTY | Identify a response from the root controller for optimal interdomain routing forwarding paths |

Finally, based on the obtained optimal intradomain and interdomain routing forwarding paths, the MDRL-TP multiagent cross-domain routing algorithm is obtained the globally optimal routing forwarding paths through the network. The following subsections introduce the designed Dueling DQN DRL algorithm, the network traffic state prediction algorithm, and the MDRL-TP multiagent cross-domain routing algorithm.

### A. Dueling DQN DRL Algorithm

The typical model for RL is a standard Markov decision process (MDP) model, which is an important branch of machine learning (ML). The MDP paradigm is used to solve the problem of optimal decision-making and to achieve the objective of maximizing utility. The basic concepts used in RL mainly consist of the state space, action space, and reward function, as illustrated in Fig. 4. First, the environment generates an initial state $s_t$. After the agent executes a corresponding action $a_t$ in accordance with the state $s_t$, a reward value $r_t$ and the next state $s_{t+1}$ are obtained. The above learning process is repeated to continuously train the agent toward greater rewards to achieve optimal decision-making.

By combining the advantages of RL in solving sequential decision problems and the advantages of DL in solving high-dimensional and complex problems, DRL based on a single agent has achieved many notable results in TE, load balancing and routing optimization, and automatic driving. However, with increasing network scale, single-agent DRL encounters problems such as heavy load and difficult convergence, hindering its ability to solve large-scale network routing optimization problems. In recent years, significant progress has been achieved in DRL for multiagent systems, and with the maturation of distributed technology, MDRL has achieved certain results in adversarial game theory, unmanned aerial vehicle (UAV) clustering, and NDC, providing critical insight into how routing optimization problems in large-scale network environments can be solved. The multiagent learning mechanism in this paper adopts the CTDE approach, which combines the advantages of centralized learning and independent learning, as shown in Fig. 5. Agents 1, 2, and 3 in the three local controllers each adopt the independent learning approach to update their own network models during training to select their own optimal strategies and generate the optimal intradomain routing forwarding paths in their own local domains. The agent in the root controller adopts the centralized learning approach, which can be expressed as:

$$(N, US, UA, P, \Upsilon, UR) \quad (1)$$

where $N$ denotes the number of agents, $US = S_{D_1} \cup S_{D_2} \cup S_{D_3}$ represents the union state space of the multiagent, $S_{D_d}$ ($d = 1,2,3$) is the state space of the $d$th network subdomain, $UA = \{ua_1, ua_2, \dots, ua_n\}$ represents the union action space of the agents ( $n = |V_{D_1} \cup V_{D_2} \cup V_{D_3}| \cdot |V_{D_1} \cup V_{D_2} \cup V_{D_3}|$ ), where $|V_{D_1} \cup V_{D_2} \cup V_{D_3}|$ is the number of SDN switches in the global network), $P$ denotes the state transfer function, $\Upsilon$ represents the discount factor that adjusts the current and future reward returns, and $UR$ denotes the union reward function, which represents the union reward value $ur_t$ obtained by the agent in the root controller when the system is in the union state $us_t \in US$ and the union action $ua_t \in UA$ is performed at time $t$.

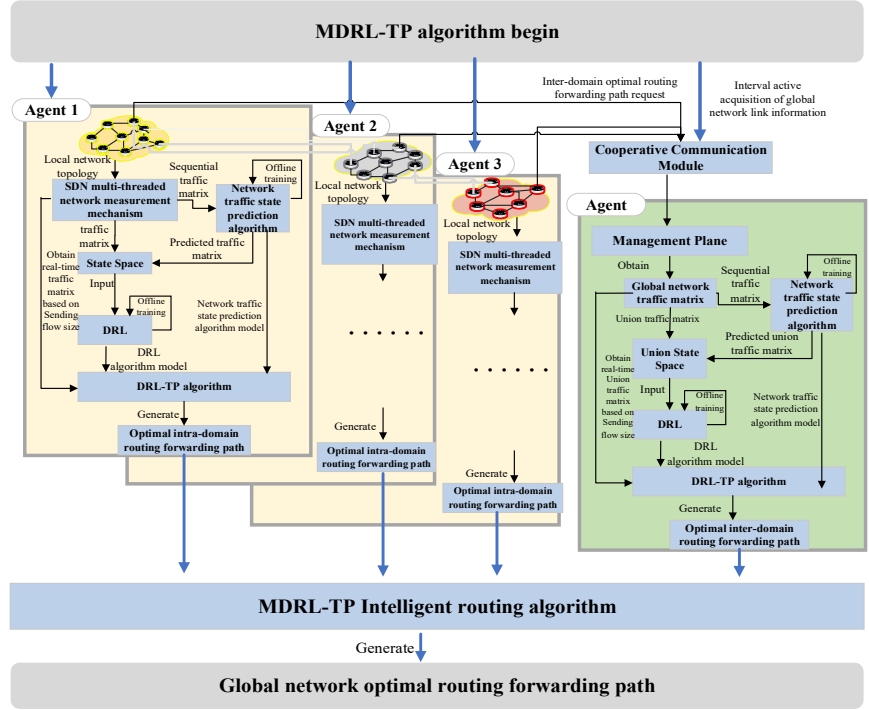

Fig. 3 Flowchart of the MDRL-TP multiagent cross-domain routing algorithm

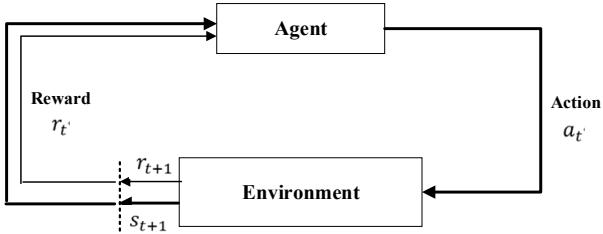

Fig. 4 RL mechanism based on a single agent

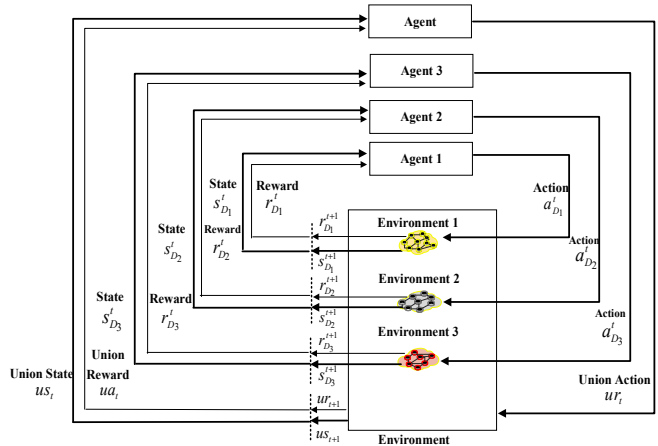

Fig. 5 RL mechanism based on multiple agents

DRL algorithm is a framework algorithm consisting mainly of composed of three elements: state, action and reward. In this paper, the design ideas for the *state space* ($S = TM$), *action space* ($A$), and *reward function* ($R$) of agents 1, 2, and 3 in the three local controllers are consistent with our previous work. Due to space limitations, please refer to [16] for the design details. Here, only the designs of the *union state space*, *union action space,* and *union reward function* for the agent in the root controller will be introduced in detail.

*Union State Space* (*US*): The union state space can be expressed as $US = UTM$, where $UTM$ is the union traffic matrix within the interval $t$, which is composed of multiple two-dimensional matrices $M_{|V_{D_1} \cup V_{D_2} \cup V_{D_3}| * |V_{D_1} \cup V_{D_2} \cup V_{D_3}|}$,

$$m_{ij} = w_1 \cdot \frac{1}{L_{bw_{ij}}} + w_2 \cdot L_{delay_{ij}} + w_3 \cdot L_{loss_{ij}}$$
$$+ w_4 \cdot L_{used_{bw_{ij}}} + w_5 \cdot L_{drops_{ij}} + w_6 \cdot L_{errors_{ij}}$$
$$, \quad i,j = 1,2,\dots,|V_{D_1} \cup V_{D_2} \cup V_{D_3}|, \quad (2)$$

where $m_{ij}$ denotes the element in the $i$th row and $j$th column of $UTM$ and is calculated from the corresponding elements of the information matrices $L_{bw}, L_{delay}, L_{loss}, L_{used\_bw}, L_{drops}$ and $L_{errors}$. Each of these information matrices, which represent the remaining bandwidth, delay, packet loss rate, used bandwidth, discarded packets, and error packets, respectively. The information matrices contain information for the links between all the switches in the network at the moment. $D_1$, $D_2$, and $D_3$ represent the three network subdomains. The weights $w_l \in [0,1]$ ($l = 1,2,\dots,6$) of the $UTM$ elements are adjustable parameters. $i$ and $j$ are represent switch names in the network, where $|V_{D_1} \cup V_{D_2} \cup V_{D_3}|$ is the total number of switches in the network. Fig. 6 shows the structure of the $UTM$.

Considering that the values in each information matrix used to form the $UTM$ are quite different, the influence of the weight factor of each information matrix on the $UTM$ cannot be objectively reflected, and the MDRL-TP multiagent cross-domain routing algorithm is difficult to converge, which affects the performance of the network model. Therefore, this paper adopts a Min-Max [40] technique to normalize the elements in the $UTM$ to the specified range $[\mu_1, \mu_2]$ to improve the convergence speed and performance of the model,

$$\overline{m_{ij}} = \mu_1 + \frac{(m_{ij} - \text{Min}(TM)) \cdot (\mu_2 - \mu_1)}{\text{Max}(TM) - \text{Min}(TM)} \quad (3)$$

where $\overline{m_{ij}}$ represents the $UTM$ element after normalization; $\text{Min}(TM)$ and $\text{Max}(TM)$ represent the minimum and maximum elements in $UTM$, respectively.

*Union Action Space* (*UA*): An action is a behavior generated by an agent in accordance with the current network state; in this paper, it represents the routing decision made by the agent. Following our previous work [16], each action $ua_t \in [0,1,\dots,k]$ in $UA$ as designed in this paper corresponds to a forwarding path selection in $us_t \in US$, which consists of $k = \lceil 0.1 \cdot \text{Max}(|V_{D_1}|, |V_{D_2}|, |V_{D_3}|) \cdot \text{Max}(|V_{D_1}|, |V_{D_2}|, |V_{D_3}|) \rceil$ candidate path matrices. The elements in each candidate path

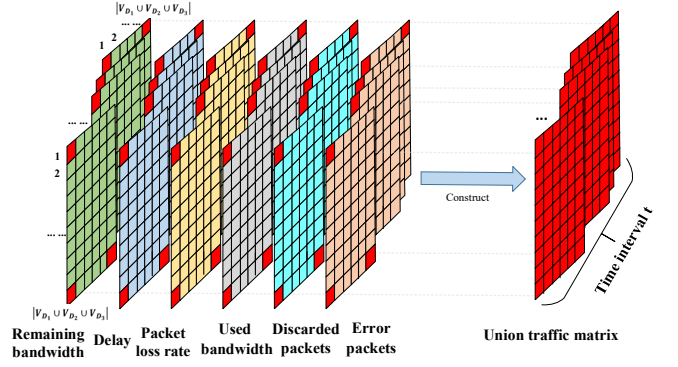

Fig. 6 Union traffic matrix structure

matrix are composed of the $path_{ij} = [i,\dots,j]$ in the global network, where $path_{ij}$ represents the path from switch $i$ to switch $j$.

*Union Reward Function* (*UR*): The return coefficient obtained by the agent after performing the action is used to guide the agent to make better action decisions in order to obtain higher rewards. Since DRL aims to continuously obtain higher rewards, in the *UR* design of multiagent cross-domain routing algorithm, the positive coefficient 1 is set for the maximization index and the negative coefficient-1 is set for the minimization index. The adjustable parameters $\varphi_l \in [0,1]$ ($l = 1,2,\dots,6$) are taken as the weight factors for constructing *UR*.

$$UR = \varphi_1 \cdot L_{bw} - \varphi_2 \cdot L_{delay} - \varphi_3 \cdot L_{loss} - \varphi_4 \cdot L_{used_{bw}}$$
$$-\varphi_5 \cdot L_{drops} - \varphi_6 \cdot L_{drops} \quad (4)$$

In this paper, Dueling DQN [41], an improved form of the DQN algorithm, is used to construct the DRL algorithm, which can better solve the overestimation of the value function generated in DQN and improve the performance of the model to find a better action strategy. Compared with the DQN, the value function of the DQN in Dueling DQN is divided into two parts. The first part is called the *value function*, which is related only to the union state $us$ and is denoted by $VF(us, \omega, \beta)$. The second part, called the *advantage function*, which depends on the values of $us$ and the $ua$, and is denoted by $AF(us, ua, \omega, \lambda)$. The complete Q function,

$$Q(us, ua, \beta, \lambda) = VF(us, \omega, \beta) + AF(us, ua, \omega, \lambda) \quad (5)$$

where $\omega$ represents the network parameters that are shared between the two parts and $\beta$ and $\lambda$ represent the network parameters that are unique to $VF$ and $AF$, respectively. Dueling DQN takes advantage of the experiential replay and greedy exploration mechanisms of the DQN algorithm. In the experiential replay mechanism, an agent stores its historical learning experiences in an experience pool $M$ in the form of $(us_t, ua_t, ur_t, us_{t+1})$ and then randomly sampling data from $M$ to provide the network model for offline training. By using experience replay mechanism, on the one hand, an agent can reuse advantageous experience samples for efficient experience sampling, thereby reducing the time overhead for obtaining experience samples. On the other hand, the use of empirical samples from different strategies reduces the correlations between data and improve the generalization ability of the algorithm. The MDRL-TP multiagent cross-domain routing algorithm also uses a decaying $\varepsilon$-greedy detection mechanism [42],

$$ua_t = \begin{cases} random.random(), & if\ x > 1 - \varepsilon \\ argmax_{ua} Q_{policy}(\Phi(us_t), ua, \theta), & otherwise \end{cases} \quad (6)$$

here, $x$ denotes the random variable between [0,1]. When $x > 1$, the agent takes exploration action, otherwise it takes exploitation action. $\varepsilon \in [0,1]$ denotes the adjustable factor, which is initialized to a value close to 1. As the number of training increases, $\varepsilon$ decreases linearly to $e_{min}$.

$$\varepsilon = e_{max} + u \cdot (e_{min} - e_{max}) \quad (7)$$

where $u = min(1, steps/totalSteps)$ represents the utilization factor, $steps$ represents the training iteration step, $totalSteps$ represents the total step size of the decaying $\varepsilon$-greedy detection mechanism, and $e_{max}$ and $e_{min}$ represent the largest and smallest exploration factor respectively.

The specific design steps of Dueling DQN deep reinforcement learning algorithm in this paper are shown in Algorithm 1, where $UTM$ consists of the interdomain traffic matrix collected by the cooperative communication module, t $TM$ within each network subdomain and the predicted traffic matrix generated by the prediction algorithm. Line 1 initializes $Q_{policy}$ and $Q_{target}$ for the same network model and experience storage pool $M$. Lines 9 to 13 update the parameters of the policy network $Q_{policy}$. Lines 14 to 16, when $steps$ reaches the update frequency $freq$, the weights and deviation of $Q_{target}$ are updated to those of $Q_{policy}$. Line 17 moves the agent to the next union state. The forwarding paths between all source-destination switches in the network are output after each iteration.

*B. Network Traffic State Prediction Algorithm*

This paper used GRU as a network traffic state prediction algorithm because GRU has fewer parameters than LSTM, which reduces the probability of overfitting of the model and accelerates the training speed of the model. Therefore, the GRU model can better meet the real-time requirements of the $UTM$ for the knowledge plane in the root controller.

The GRU-based network traffic state prediction algorithm is shown in Algorithm 2. Line 2 generates $UTM_{input}$ and $UTM_{target}$ as the input union matrix and target union matrix, respectively, of the GRU model after the time-series operation. Lines 4 to 8 train the GRU model, where Line 6 generates an output time-series union traffic matrix and the model's next input parameter $hidden_{output}$. After training, use the trained model outputs the predicted union traffic matrix, which is used to construct $UTM$.

---

**Algorithm 1** Dueling DQN DRL algorithm

**Input**:
  Learning rate: $lr$
  Sampling size: $batch$
  Discount factor: $\gamma$
  Weight factors: $\varphi_l \in [0,1], l = 1,2,...,6$
  Attenuation parameter: $\varepsilon$
  Attenuation rate: $decay$
  Target network update frequency: $freq$
  Total number of training episodes: $episodes$
  Union traffic matrix: $UTM$
**Output**:
  Forwarding paths between all source-destination switches in the network
1  Initialize $Q_{policy}$ and $Q_{target}$ network weights $\theta$, experience pool $M$
2  **for** $episodes \leftarrow 1$ to n do:
3    The agent obtains the initial union state $us_t$
4    **while** next union state $us_{t+1}$ is not final state do:
5      Update attenuation parameter $\varepsilon = \varepsilon - (steps \cdot decay)$
6      Obtain $ua_t$ under $us_t$ according to Equation (6)
7      Obtain the current $ur_t \leftarrow UR(us_t, ua_t)$ according to Equation (3)
8      Store experience $(us_t, ua_t, ur_t, us_{t+1})$ in $M$
9      **if** $len(M) \geq batch$ then:
10       Randomly sampling $batch$ data from $M$
11       Obtain $p\_value$ and $t\_value$ according to Equation (5):
         $p\_value = Q_{policy}(us_t, ua_t, \theta)$
         $t\_value = \begin{cases} ur_t, & if\ next\ union\ state\ is\ final\ state \\ ur_t + \gamma \cdot max_{ua'} Q_{target}(us_{t+1}, ua'; \theta), & otherwise \end{cases}$
12       Update the $Q_{policy}$ network weight parameters $\theta$ by executing gradient descent using $loss = (t\_value - p\_value)^2$
13     **end if**
14     **if** $steps\ \%\ freq\ == 0$ then:
15       Update the $Q_{target}$ network model parameters $\theta^{Q_{target}} \leftarrow \tau \cdot \theta^{Q_{policy}} + (1-\tau) \cdot \theta^{Q_{target}}$
16     **end if**
17     $us_t \leftarrow us_{t+1}$
18   **end while**
19 **end for**

---

**Algorithm 2** Network traffic state prediction algorithm

**Input**:
  Learning rate: $lr$
  Sampling size: $batch$
  Input layer dimension: $input\_dim$
  Hidden layer dimension: $hidden\_dim$
  Output layer dimension: $output\_dim$
  Sequential period: $seq$
  Target network update frequency: $freq$
  Total number of training episodes: $episodes$
  Union traffic matrix: $UTM$
**Output**:
  Predicted union traffic matrix
1  Initialize GRU network weights $\theta$
2  Execute the time-series operation on $UTM$ to obtain two time-series union traffic matrices
     input time-series union traffic matrix $UTM_{input} \leftarrow UTM$
     target time-series union traffic matrix $UTM_{target} \leftarrow UTM$
3  **for** $episodes \leftarrow 1$ to n do:
4    Initialize $hidden$
5    **for** $t \leftarrow 0$ to $len(UTM_{input})$ do:
6      $UTM_{output}^{t+seq+1}, hidden_{output} \leftarrow GRU(UTM_{output}^{t,t+seq}, hidden)$
7      Update the GRU network weight parameters $\theta$ by executing gradient descent using $loss = (UTM_{output}^{t+seq+1} - UTM_{target}^{t+seq+1})^2$
8      $hidden \leftarrow hidden_{output}$
9    **end for**
10 **end for**

## C. MDRL-TP Multiagent Cross-Domain Routing Algorithm

Algorithm 3 is an improved form of DRL-TP algorithm proposed in our previous work [16], which generates the optimal intradomain routing forwarding path in the local controller and the optimal interdomain routing forwarding path in the root controller according to the type of sending flow. The MDRL-TP multiagent cross-domain routing algorithm proposed in the current work is mainly composed of the DRL-TP intelligent routing algorithm deployed on multiple agents, as shown in Algorithm 4. Line 2 represents the size of each sent network data flow. Lines 3 to 6 obtain the optimal intradomain routing paths $intra\_optimal\_path_{D_1}$, $intra\_optimal\_path_{D_2}$ and $intra\_optimal\_path_{D_3}$ of the three network subdomains. On Line 7, the root controller obtains $UTM$ for the global network through the cooperative communication module via socket technology. On Line 8, the root controller generates the optimal interdomain routing forwarding paths $inter\_optimal\_path$ based on the interdomain routing forwarding requests from each network subdomain. On Line 9, based on the obtained optimal routing paths in each network subdomain and the optimal interdomain routing and forwarding paths between network subdomains, the optimal routing paths $all\_optimal\_path$ from the source-destination switch nodes in the global network are generated by the MDRL-TP multiagent cross-domain routing algorithm.

---

**Algorithm 3** DRL-TP single-agent routing algorithm

**Input**:
   Dueling DQN algorithm
   GRU-based prediction algorithm
   Sent flow size: $bw\_list$

**Output**:
   Optimal forwarding paths for all source-destination switch in each network subdomain

1 Load the Dueling DQN and GRU models to build the DRL-TP model
2 **for** $bw \leftarrow 0$ to $bw\_list$ **do**:
3   **if** $bw$ in the local controller:
4     $intra\_optimal\_path \leftarrow$ DRL-TP($TM$)
5   **else if** $bw$ in the root controller:
6     $inter\_optimal\_path \leftarrow$ DRL-TP($UTM$)
7   **end if**
8 **end for**

---

**Algorithm 4** MDRL-TP multiagent cross-domain routing algorithm

**Input**:
   DRL-TP algorithm
   Sent flow size: $bw\_list$

**Output**:
   Optimal forwarding paths for all source-destination switch in the global network

1 Load DRL-TP routing models and build the MDRL-TP multiagent cross-domain routing model
2 **for** $bw \leftarrow 1$ to $bw\_list$ **do**:
3   **for** $D_d \leftarrow$ 1 to 3 **do**:
4     Obtain the intradomain traffic matrix $TM_{D_d}$ in the subdomain of the network
5     $intra\_optimal\_path_{D_i} \leftarrow$ DRL-TP($TM_{D_d}$)
6   **end for**
7   The root controller obtains $UTM$ through the cooperative communication module
8   The root controller generates the optimal interdomain routing forwarding paths $inter\_optimal\_path \leftarrow$DRL-TP($UTM$)
9   $all\_optimal\_path \leftarrow$MDRL-TP($intra\_optimal\_path_{D_1 \cup D_2 \cup D_3} \cup inter\_optimal\_path$)
10 **end for**

---

## V. EXPERIMENTAL ANALYSIS

### A. Experimental Environment

In our wok, four Ubuntu 16.04 systems with 4-core processors and memory sizes of 2 GB are used to build a hierarchical multicontroller SDN architecture, three of which are used as local controllers, while the remaining one is the root controller. The four Ubuntu 16.04 systems are equipped with Mininet 2.3.0 [43] and Ryu 4.34 [44] to build the SDN environment. The topology in the SDN network environment is generated by Mininet, and the SDN controller uses Ryu based on Python open source, and uses Iperf [45] tool to create and send data streams. As shown in Fig. 7, the modified New York City Center [46] network topology consists of 39 switch nodes，which are divided into three network subdomains. Each node represents a switch that supports the OpenFlow 1.3 protocol, and each switch is mounted under a host. Considering the limitations of the experimental hardware devices and the requirements of new heterogeneous networks, the link bandwidth between all switches is randomly set to [1 Mbit–10 Mbit].

### B. Prediction Algorithm Performance and Experimental Parameter Analysis

As shown in Fig. 8, by using the GRU-based network traffic state prediction algorithm, an agent can learn to obtain a higher reward value. The reason is that the GRU-based prediction algorithm can monitor hidden network traffic states in a large-scale SDN network under multicontroller management, which are difficult to obtain based solely on the multithreaded SDN network measurement mechanism and cooperative communication module. Moreover, the GRU-based prediction algorithm can predict the future trend of the network traffic state, allowing the MDRL-TP multiagent cross-domain routing algorithm to obtain more state space and explore a better action strategy in order to obtain higher rewards. These results effectively verify that the use of the GRU prediction algorithm can improve the performance of the MDRL-TP multiagent cross-domain routing algorithm. In our previous work [16], we performed a detailed parameter comparison experiment focusing on the optimal weight factors for the traffic matrix and the reward function and the number of network link indices that make up the traffic matrix. Considering the similarity of the data distribution of the union traffic matrix, this paper follows the same design idea as the previous work [16] and uses [0.6,0.3,0.1,0.1,0.1,0.1] and [0.5, -0.4, -0.3, -0.3 0.3, -0.3, -0.3, -0.3] as the weight factors for $UTM$ and $UR$. The $UTM$ elements are calculated by combining six network link indicators: remaining bandwidth, delay, packet loss rate, used bandwidth, discarded packets, and error packets. Fig. 9, Fig. 10, and Fig. 11 compare the results for different values of the weight factors for $UTM$ and $UR$ and

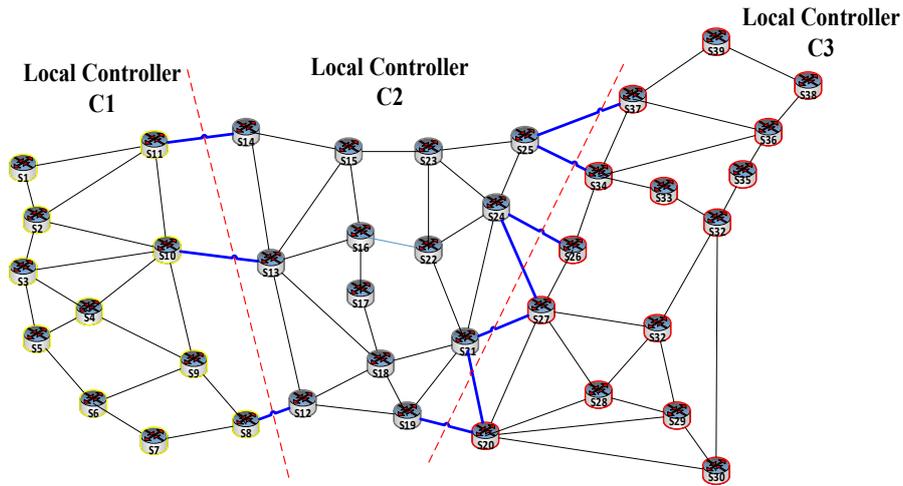

Fig. 7 Modified New York City Center network

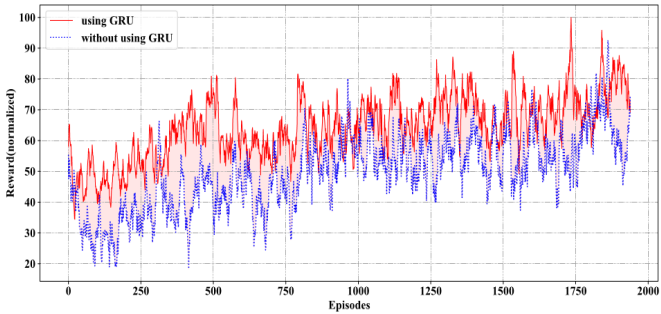

(a) Agent in root controller

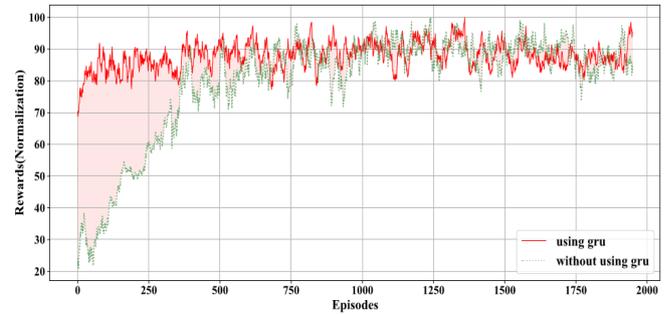

(b) Agent 1 in local controller C1

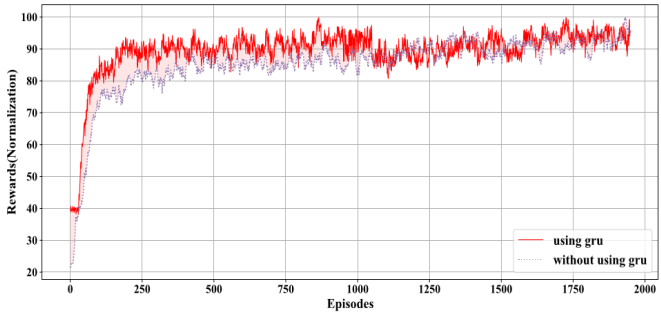

(c) Agent 2 in local controller C2

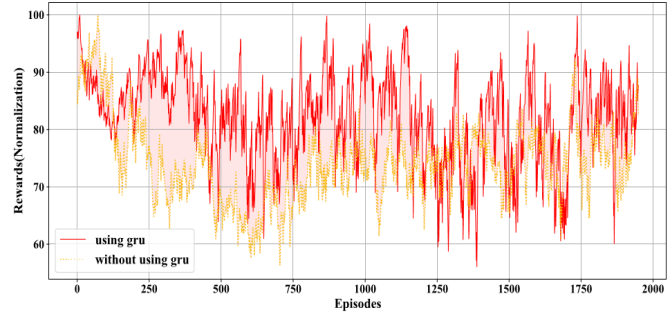

(d) Agent 3 in local controller C3

Fig. 8 Comparison of training results with and without the GRU-based network prediction algorithm

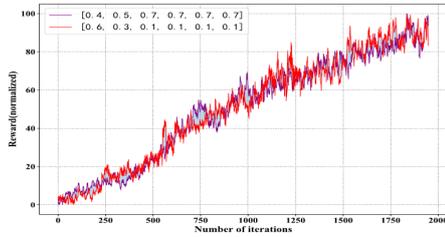

Fig. 9 Comparison of weight factors for the *UTM*

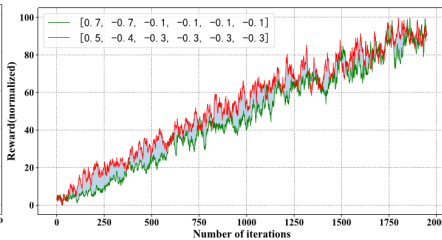

Fig. 10 Comparison of weight factors for the *UR*

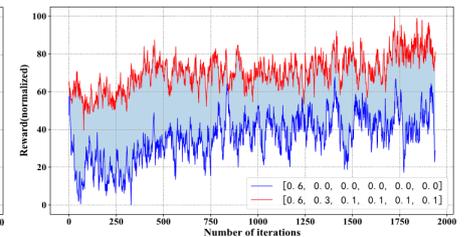

Fig. 11 Comparison of link indicators composing the *UTM*

><

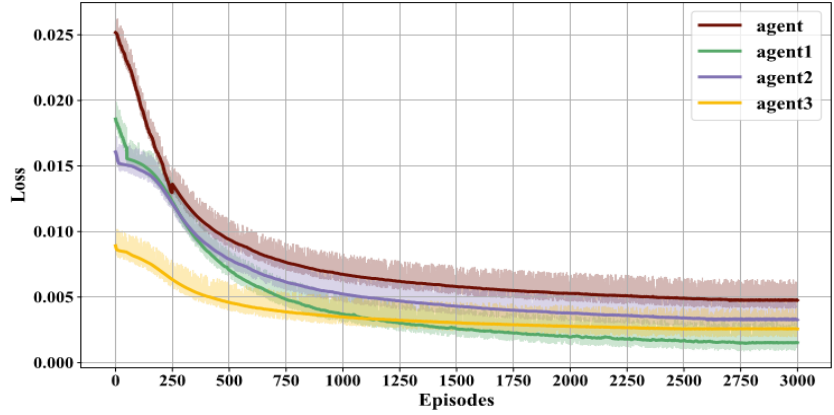

Fig. 12 Loss curves of the network traffic state prediction algorithm

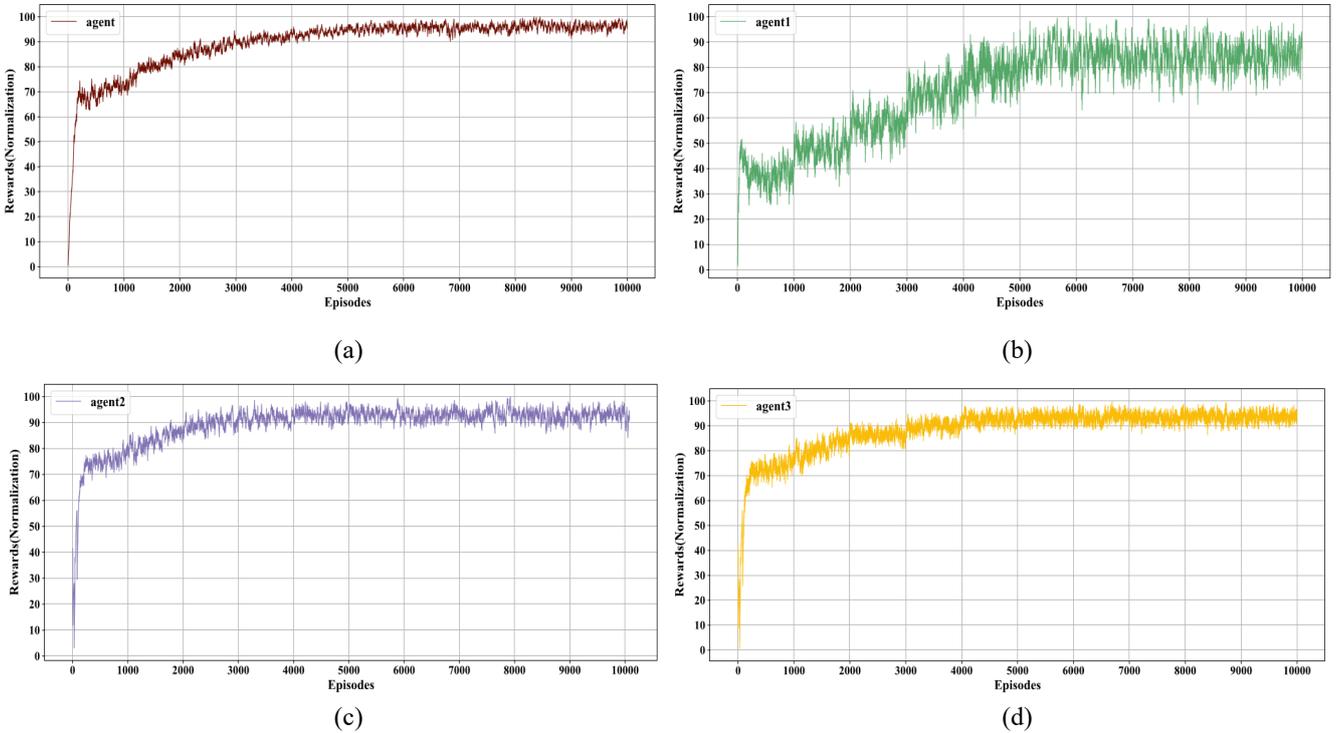

Fig. 13 Reward return curves

different numbers of network link indicators comprising $UTM$, where the values listed in the legends correspond, from left to right, to $L_{bw}$, $L_{delay}$, $L_{loss}$, $L_{used\_bw}$, $L_{drops}$ and $L_{errors}$.

The following conclusions can be drawn from Fig. 12 and Fig. 13: 1) Compared with agents 1, 2, and 3 in the local controllers, the agent in the root controller shows higher loss values of the GRU model and a higher number of iterations before reward convergence of the DRL model. This is because $UTM$ is constructed based on the union state space in the root controller, and thus, the dimensions of $UTM$ are greater than those of the in-domain traffic matrices considered by the three local controllers. 2) The reward values obtained by agent 1 are lower than those obtained by other agents, and the convergence degree of this agent is also worse. The main reason is that we wish to verify whether the MDRL-TP multiagent cross-domain routing algorithm can guarantee the global network performance when a certain network subdomain is congested. Therefore, the link bandwidth in the network subdomain under the management of controller C1 is approximately limited to 5 Mbit. Consequently, when the number of sent data flows increases, the network subdomain managed by C1 is more likely to be in a congested state, resulting in poorer performance of the obtained intradomain traffic matrix and affecting the convergence of the DRL model.

## C. Comparative Experiments and Results

In this paper, the Dijkstra and open shortest path first (OSPF) routing algorithms are considered for comparison with the proposed MDRL-TP multiagent cross-domain routing algorithm. The implementation ideas of the OSPF and Dijkstra algorithms are introduced below.

*Dijkstra Routing Algorithm:* When the SDN network topology is constructed, by setting $W$ to 1 as the link weight value between switches, the forwarding path of each source-destination switch node is obtained based on the shortest-hop.

*OSPF Routing Algorithm:* The intradomain link delays in each network subdomain are obtained through the multithreaded SDN network measurement mechanism, and the interdomain link delays between network subdomains are obtained by sending detection packets $detect\_pkt$. Then, in accordance with the network link delays, the paths between all switch nodes are obtained, and finally, the shortest-hop paths at the current time are selected as the routing and forwarding paths.

In this paper, network throughput, delay and packet loss rate are designed to compare the network performance based on three routing algorithms. In order to better verify the effectiveness of MDRL-TP cross-domain intelligent routing algorithm, the intradomain network indicators in each subdomain and the global network indicators are compared experimentally. The intradomain indicators are expressed as:

$$throughtput_{D_{d_{e_{ij}}}} = \frac{tx_{bytes}(e_{ij})}{bw_t(e_{ij}) \cdot |\Delta t|}, \quad d = 1,2,3 \quad (8)$$

$$throughtput_{D_{1_{e_{ij}}} \cup D_{2_{e_{ij}}} \cup D_{3_{e_{ij}}}} = \frac{tx\_bytes(e_{ij})}{bw_t(e_{ij}) \cdot |\Delta t|} \quad (9)$$

here, $i$ and $j$ are indices representing particular switch names. $D_d$ ( $d = 1,2,3$ ) represents the $d$ th network subdomain. $throughput_{D_{d_{e_{ij}}}}$ and $throughput_{D_{1_{e_{ij}}} \cup D_{2_{e_{ij}}} \cup D_{3_{e_{ij}}}}$ represent the throughput of link $e_{ij}$ in network subdomain $D_d$ and the throughput of link $e_{ij}$ in the whole network, respectively, in time interval $\Delta t$. $tx\_bytes(e_{ij})$ represents the number of bytes sent from switch $i$ to $j$ within $\Delta t$, and $bw_t(e_{ij})$ represents the remaining bandwidth of link $e_{ij}$ in $\Delta t$.

The intradomain link delay in each network subdomain can be obtained by the SDN network link delay measurement mechanism [47]. As shown in Fig. 14, $T_1$ and $T_2$ are each obtained through LLDP. Then, the SDN controller obtains the time $T_a$ and $T_b$ by sending the echo message to the switches *SA* and *SB*, and the intradomain network link delay is calculated according to (10). The measurement method for the interdomain network link delay is illustrated in Fig. 15. Switch S3 under controller C1 first obtains $T_{inter1}$ by sending a detection packet $detect\_pkt$ to switch S6 under controller C3. Then, S6 sends $detect\_pkt$ to S3 to obtain $T_{inter2}$, and the interdomain network link delay is calculated according to (11). Finally, the global network link delay is expressed as:

$$delay_{D_{d_{e_{ij}}}} = \frac{T_1 + T_2 + T_3 + T_4}{2}, \quad d = 1,2,3 \quad (10)$$

$$delay_{e_{ij}} = \frac{T_{inter1} + T_{inter2}}{2} \quad (11)$$

$$delay_{D_{1_{e_{ij}}} \cup D_{2_{e_{ij}}} \cup D_{3_{e_{ij}}}} = \begin{cases} delay_{D_{d_{e_{ij}}}}, & if\ e_{ij} \in D_d, d = 1,2,3 \\ delay_{e_{ij}}, & otherwise \end{cases} \quad (12)$$

here, when link $e_{ij}$ is an intradomain link, $delay_{D_{d_{e_{ij}}}}$ is used to represent the global network link delay, and when link $e_{ij}$ is an interdomain link, $delay_{e_{ij}}$ represents the global network link delay. Similarly, the network link packet loss rate and the global network link packet loss rate are expressed as:

$$loss_{D_{d_{e_{ij}}}} = \frac{tx\_pkts(e_{ij}) - rx\_pkts(e_{ij})}{tx\_pkts(e_{ij})}, d = 1,2,3 \quad (13)$$

$$loss_{D_{1_{e_{ij}}} \cup D_{2_{e_{ij}}} \cup D_{3_{e_{ij}}}} = \frac{tx\_pkts(e_{ij}) - rx\_pkts(e_{ij})}{tx\_pkts(e_{ij})} \quad (14)$$

here, $i$ and $j$ are indices representing particular switch names. $D_d$ ( $d = 1,2,3$ ) represents the $d$ th network subdomain, $loss_{D_{d_{e_{ij}}}}$ represents the packet loss rate of link $e_{ij}$ in network subdomain $D_d$ within time interval $\Delta t$, $loss_{D_{1_{e_{ij}}} \cup D_{2_{e_{ij}}} \cup D_{3_{e_{ij}}}}$ represents the packet loss rate of link $e_{ij}$ in the global network within $\Delta t$. $tx\_pkts(e_{ij})$ represents the number of packets sent from switch $i$ to $j$ within $\Delta t$, and $rx\_pkts(e_{ij})$ represents the number of packets received by switch $i$ from $j$ within $\Delta t$. Considering that the performance of an SDN network may exhibit some fluctuations during long-term operation, the three evaluation indexes based on single acquisition of network throughput, delay and packet loss rate cannot truly reflect the network performance at the current time. Therefore, to reduce the impact of network fluctuations, the averaging method is applied in this paper to compare the impacts of the three routing algorithms on network performance.

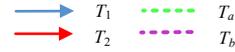
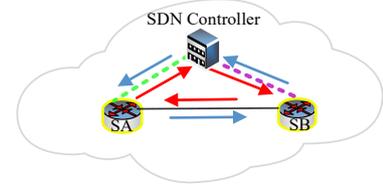

Fig. 14 Intradomain network link delay measurement method

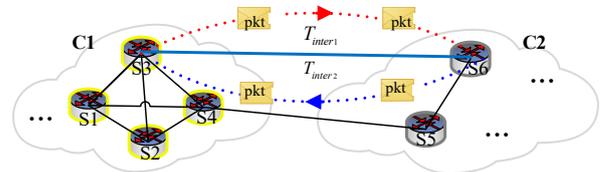

Fig. 15 Interdomain network link delay measurement method

First, the intradomain network performance indicators in each network subdomain are compared,

$$domain\_avg\_throughput_{D_d}$$
$$= \frac{\sum_i^{|V_d|} \sum_j^{|V_d|} throughput_{D_{d_{e_{ij}}}}}{|V_{D_d}| \cdot |V_{D_d}|}$$

$$domain\_avg\_delay_{D_d} = \frac{\sum_i^{|V_d|} \sum_j^{|V_d|} delay_{D_{d_{e_{ij}}}}}{|V_{D_d}| \cdot |V_{D_d}|}$$

$$domain\_avg\_loss_{D_d} = \frac{\sum_i^{|V_d|} \sum_j^{|V_d|} loss_{D_{d_{e_{ij}}}}}{|V_{D_d}| \cdot |V_{D_d}|}$$

$$i, j \in V_{D_d}, d = 1,2,3 \quad (15)$$

here, $domain\_avg\_throughput_{D_d}$, $domain\_avg\_delay_{D_d}$ and $domain\_avg\_loss_{D_l}$ denote the average throughput, average delay, and average packet loss rate, respectively, in the current subdomain $D_d$. Some conclusions can be seen from Fig. 16, Fig. 17, and Fig. 18: 1) Compared with the network subdomains managed by local controllers C2 and C3, the network subdomain managed by local controller C1 shows significantly lower intradomain average throughput. In this experiment, the link bandwidths throughout the global network topology are randomly set to values in the range of [1 Mbit–10 Mbit] to better verify performance of the MDRL-TP multiagent cross-domain routing algorithm proposed in this paper; however, the link bandwidths in the network subdomain under controller C1 are mostly limited to 5 Mbit. Thereby, when the size of sending flow exceeds 5 Mbit/s, obvious congestion occurs in the subdomain managed by C1, causing the intradomain average delay in Fig. 16 (b) and the intradomain average packet loss rate in Fig. 16 (c) to increase rapidly and thus affecting the intradomain average throughput. 2) The intradomain average throughputs, delays and packet loss rates in the network subdomains managed by local controllers C2 and C3 are similar, but the intradomain network performance indicators of the network subdomain managed by C3 are somewhat better than those under C2. On the one hand, the network link bandwidth under C3 is generally greater than that under C2; on the other hand, local controller C2, as an intermediate controller, is more greatly impacted by C1 than C3 when local controller C1 is congested during interdomain information exchange. Therefore, the network subdomain managed by C3 shows better intradomain network performance indicators. 3) When the network state is either normal or congested, compared with the Dijkstra and OSPF routing algorithms, the MDRL-TP multiagent cross-domain routing algorithm in this paper can improve the performance in terms of the intradomain network indicators, thus verifying the effectiveness of this intelligent routing algorithm.

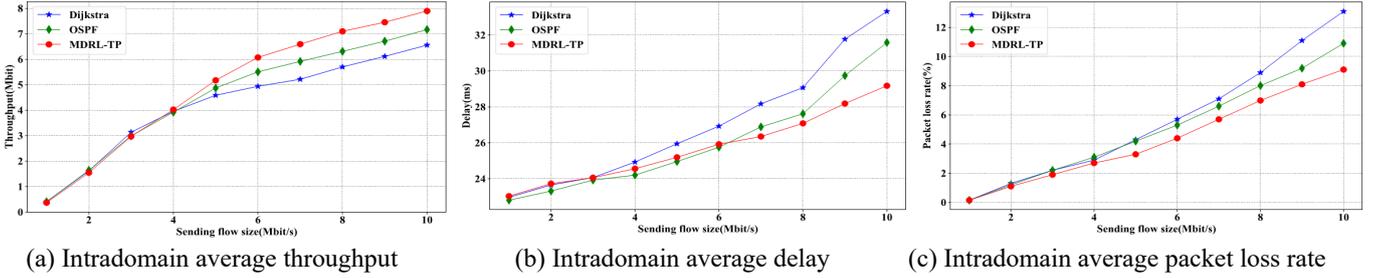

(a) Intradomain average throughput    (b) Intradomain average delay    (c) Intradomain average packet loss rate

Fig. 16 Comparison of network performance indicators under local controller C1

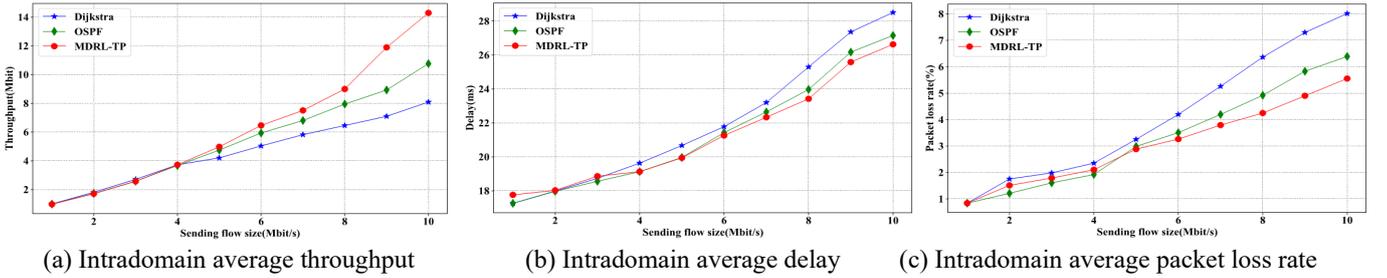

(a) Intradomain average throughput    (b) Intradomain average delay    (c) Intradomain average packet loss rate

Fig. 17 Comparison of network performance indicators under local controller C2

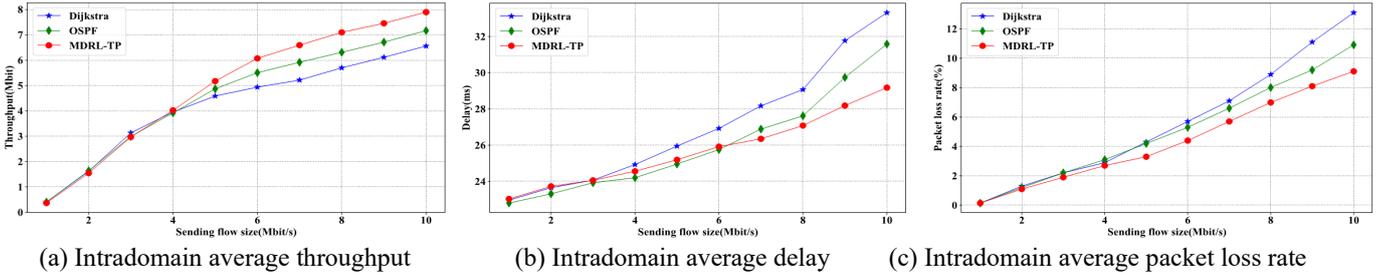

(a) Intradomain average throughput    (b) Intradomain average delay    (c) Intradomain average packet loss rate

Fig. 18 Comparison of network performance indicators under local controller C3

Next, the global network performance indicators are compared,

$$all\_avg\_throughput = \frac{\sum_{i}^{|V_{D_1} \cup V_{D_2} \cup V_{D_3}|} \sum_{j}^{|V_{D_1} \cup V_{D_2} \cup V_{D_3}|} throughput_{D_{1e_{ij}} \cup D_{2e_{ij}} \cup D_{3e_{ij}}}}{|V_{D_1} \cup V_{D_2} \cup V_{D_3}| \cdot |V_{D_1} \cup V_{D_2} \cup V_{D_3}|}$$

$$all\_avg\_delay = \frac{\sum_{i}^{|V_{D_1} \cup V_{D_2} \cup V_{D_3}|} \sum_{j}^{|V_{D_1} \cup V_{D_2} \cup V_{D_3}|} delay_{D_{1e_{ij}} \cup D_{2e_{ij}} \cup D_{3e_{ij}}}}{|V_{D_1} \cup V_{D_2} \cup V_{D_3}| \cdot |V_{D_1} \cup V_{D_2} \cup V_{D_3}|}$$

$$all\_avg\_loss = \frac{\sum_{i}^{|V_{D_1} \cup V_{D_2} \cup V_{D_3}|} \sum_{j}^{|V_{D_1} \cup V_{D_2} \cup V_{D_3}|} loss_{D_{1e_{ij}} \cup D_{2e_{ij}} \cup D_{3e_{ij}}}}{|V_{D_1} \cup V_{D_2} \cup V_{D_3}| \cdot |V_{D_1} \cup V_{D_2} \cup V_{D_3}|}$$

$$i, j \in |V_{D_1} \cup V_{D_2} \cup V_{D_3}| \quad (16)$$

here, $all\_avg\_throughput$, $all\_avg\_delay$ and $all\_avg\_loss$ denote the average throughput, average delay and average packet loss rate, respectively, of the whole network, and $i$ and $j$ are representing particular switch names. We can draw the following conclusions from Fig. 19, Fig. 20, and Fig. 21: 1) When the sent flow size reaches 5 Mbit/s, the average throughput, delay, and packet loss rate in the global network change markedly. From this behavior, it can be inferred that the network began to appear congestion. However, compared with the Dijkstra and OSPF routing algorithms, the MDRL-TP multiagent cross-domain routing algorithm can still ensure the global performance of the network. 2) The global network performance is easily affected by the performance of one network subdomain. Therefore, in comparison with Fig. 17 (a) and Fig. 18 (a), the average throughput values shown in Fig. 19 are lower because the network subdomain managed by local controller C1 is affected by the restricted link bandwidth. When the sent flow size is greater than 5 Mbit/s, this subdomain of the network is seriously congested, which greatly affects the performance of the domain, and also affects the performance of the global network. 3) Even if the network performance in one subdomain is poor, the proposed MDRL-TP multiagent cross-domain routing algorithm can still ensure the global network performance, thus verifying its effectiveness.

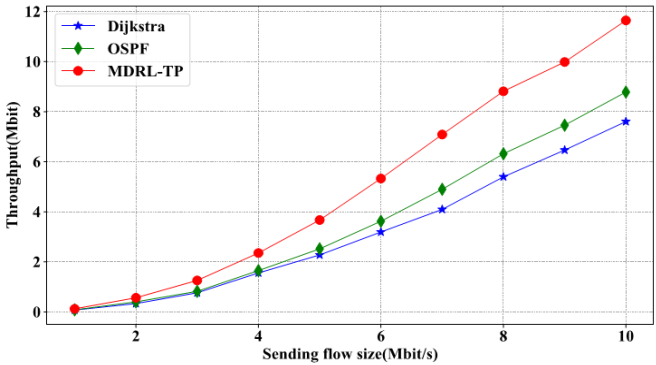

Fig. 19 Global average throughput

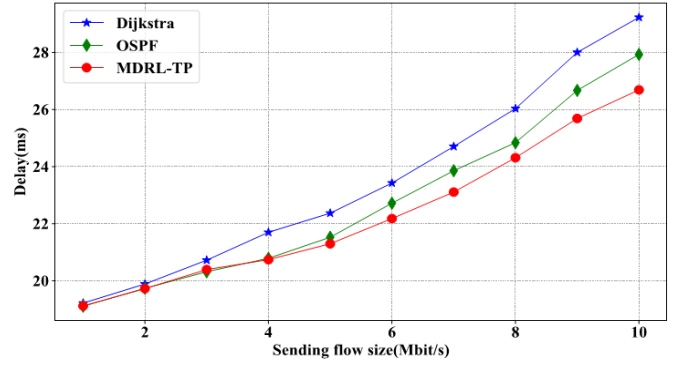

Fig. 20 Global average delay

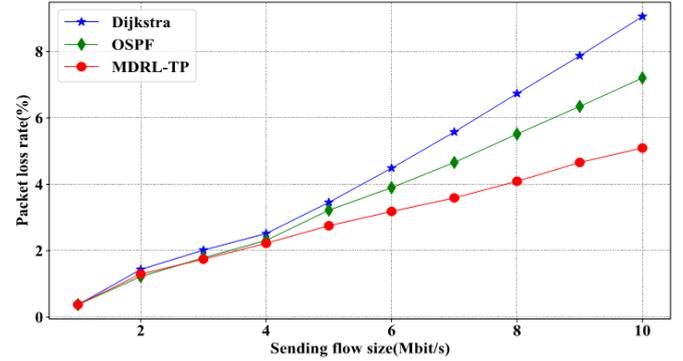

Fig. 21 Global average packet loss rate

## VI. CONCLUSION

With the increasing scale of SDN networks, intelligent routing algorithms based on a single controller will encounter problems of excessive load, a single point of failure, and message accumulation for large-scale networks. To solve the routing optimization problem for large-scale SDN networks, a feasible method is to divide one large-scale network into multiple smaller network subdomains and use multiple SDN controllers for management. To this end, we design a cross-domain intelligent SDN routing method based on a proposed MDRL algorithm. Each local controller generates the optimal intradomain routing forwarding paths for its own network subdomain, and the optimal interdomain routing forwarding paths are generated by the root controller in response to requests sent through the cooperative communication module via socket technology; thus, the optimal forwarding paths for the global network are finally generated. Compared with the Dijkstra and OSPF routing algorithms, the proposed MDRL-TP multiagent cross-domain routing algorithm can better guarantee the intradomain network performance in each network subdomain, effectively improve the average throughput in the global network, and reduce the average delay and packet loss rate in the global network. Thus, the proposed algorithm has practical significance for addressing the routing optimization problem in large-scale SDN networks. The influence of the SDN multicontroller deployment location and the number of network subdomains on large-scale SDN routing optimization will be the direction of future research.

**Miao Ye** received his B.S degree in theory physics from Beijing Normal University in 2000 and his Ph.D. degree from School of Computer Science and Technology from Xidian University in 2006. He is currently a full professor and Ph.D. supervisor at Guilin University of Electronic Technology. His research interests include software defined networks, edge computing and edge storage, wireless sensor networks, deep learning.

**Linqiang Huang** was born in 1998. He is a PhD student in the School of Computer Science and Information Security, Guilin University of Electronic Technology. His main research interests include reinforcement learning and software defined networking.

**Xiaofang Deng** received the B.Eng. and M.Eng. degrees in communication engineering from the Guilin University of Electronic Technology (GUET), China, in 1998 and 2005, and the Ph.D. degree from the South China University of Technology (SCUT), Guangzhou, China, in 2016. She was a Visiting Scholar with Coventry University, in 2017. She is currently an Associate Professor with the School of Information and Communication Engineering, GUET. Her research interests include cognitive networks, network economy, and information sharing.

**Yong Wang** was born in 1964. He received his Ph.D. degree from East China University of Science and Technology, Shanghai, China, in 2005. He is currently a full professor and Ph.D. supervisor at Guilin University of Electronic Technology. His main research interests are cloud computing, distributed storage systems, software defined networks and information security.

**Qiuxiang Jiang** was born in 1978. She received her Master's degree from Guilin University of Technology, Guilin, China, in 2005. She is a senior engineer at Guilin University of Electronic Technology. Her main research interests are software defined networks, wireless sensor network and cloud computing.

**Hongbing Qiu** was born in 1963, Ph.D., professor of Guilin University of Electronic Science and Technology, doctoral supervisor of Xidian University and Guilin University of Electronic Science and Technology, visiting researcher at University of Minnesota，Twin Cities in 2012, communication theory of China Academy of Communications Member of the Signal Processing Committee, Director of the Key Laboratory of Cognitive Radio and Information Processing of the Ministry of Education, member of the Broadband Wireless IP Standard Working Group, etc. His main research directions are wireless communication, ultra-wideband communication, wireless sensor networks and software defined networks.

**Peng Wen** was born in 1994. He received his B.S. degree in Mathematics and Applied Mathematics from Yangzhou University, Jiangsu, PR China. He received his M.S. in Probability Theory and Mathematical Statistics from Guangxi Normal University, Guilin, Guangxi, PR China. Now, he is a PhD student in School of Information and Communication, Guilin University of Electronic Technology. His research interests are broadly in the areas of reinforcement learning, software defined networking, stochastic processes and queueing models for communication networks.